\documentclass[aps,prc,superscriptaddress,nofootinbib,showpacs,floatfix,singlecolumn]{revtex4}
\usepackage{graphicx} 
\usepackage{float} 
\usepackage{amssymb}
\usepackage{url}
\usepackage[colorlinks,linkcolor=blue,citecolor=blue,filecolor=black,urlcolor=blue]{hyperref}
\usepackage{cancel}
\usepackage{MnSymbol}
\usepackage{bm}

\newcommand{\lr}[1]{\left\langle #1\right\rangle}

\newcommand{\etaref}{\eta_{\mathrm{r}}}

\usepackage{color}
\usepackage{harpoon}
\definecolor{my}{rgb}{1, 0, 0}

\begin{document} 
\title{Observables for longitudinal flow correlations in heavy-ion collisions}
\newcommand{\sunysb}{Department of Chemistry, Stony Brook University, Stony Brook, NY 11794, USA}
\newcommand{\bnl}{Physics Department, Brookhaven National Laboratory, Upton, NY 11796, USA}{
\newcommand{\shanghai}{Shanghai Institute of Applied Physics, Chinese Academy of Sciences, Shanghai 201800, China}
\newcommand{\beijing}{University of Chinese Academy of Sciences, Beijing 100049, China}
\author{Jiangyong Jia}\email[Correspond to\ ]{jjia@bnl.gov}\affiliation{\sunysb}\affiliation{\bnl}
\author{Peng Huo}\affiliation{\sunysb}
\author{Guoliang Ma}\affiliation{\shanghai}
\author{Maowu Nie}\affiliation{\sunysb}\affiliation{\shanghai}\affiliation{\beijing}

\begin{abstract}
We propose several new observables/correlators, based on correlations between two or more subevents separated in pseudorapidity $\eta$, to study the longitudinal flow fluctuations. We show that these observables are sensitive to the event-by-event fluctuations, as a function of $\eta$, of the initial condition as well as the non-linear mode-mixing effects. Experimental measurement of these observables shall provide important new constraints on the boost-variant event-by-event initial conditions required by all 3+1-dimensional viscous hydrodynamics models. 
\end{abstract}
 \maketitle
\section{Introduction}
Multi-particle azimuthal correlations are a powerful tool to study the collective behavior of the quark-gluon plasma created in relativistic heavy ion collisions at RHIC and the LHC~\cite{Borghini:2001vi,Bilandzic:2010jr,Bilandzic:2013kga}. Studies of the particle correlation in the transverse plane revealed strong harmonic modulation of the particle densities in the azimuthal angle: $dN/d\phi\propto1+2\sum_{n=1}^{\infty}v_{n}\cos n(\phi-\Phi_{n})$, where $v_n$ and $\Phi_n$ represent the magnitude and phase (or event-plane angle) of the $n^{\mathrm{th}}$-order harmonic flow. Extensive measurements in the last decade have provided detailed differential information on the event-averaged $v_n$ coefficients, as well as the event-by-event fluctuations of the $v_n$ and correlations between $v_n$ and $\Phi_n$~\cite{Luzum:2013yya,Jia:2014jca}. Comparison of these measurements with event-by-event hydrodynamic model calculations has provided important constraints on the properties of the medium and transverse density fluctuations in the initial state~\cite{Heinz:2013th,Gale:2013da}.

Most earlier measurements and model calculations assumed that the initial condition and space-time evolution of the matter is boost invariant in the longitudinal direction. However, recent studies of the two-particle correlations as a function of pseudorapidity $\eta$ revealed strong event-by-event fluctuations of the flow magnitude and phase between two well-separated pseudorapidities $\eta_1$ and $\eta_2$, i.e. $v_n(\eta_1)\neq v_n(\eta_2)$ (forward-backward asymmetry) and $\Phi_{n}(\eta_1)\neq \Phi_{n}(\eta_2)$ (event-plane twist)~\cite{Bozek:2010vz,Jia:2014ysa,Huo:2013qma}, collectively referred to as ``flow decorrelations''. Measurement of a new observable by CMS collaboration~\cite{Khachatryan:2015oea} shows that the decorrelation effects vary linearly with the $\eta_1-\eta_2$, and have a strong centrality dependence for elliptic flow $v_2$ but very weak dependence for $v_3$ and $v_4$. This measurement has provided important constraint on the initial condition in longitudinal direction, which is a necessary ingredient for developing full 3+1-dimensional viscous hydrodynamic models~\cite{Bozek:2015bna,Pang:2015zrq,Ma:2016fve,Schenke:2016ksl,Ke:2016jrd}.

In this paper, we proposed a set of new observables/correlators, based on correlations between $\eta$-separated subevents~\cite{jjia}, that can be used to further clarify the longitudinal flow fluctuations. The previous observable used by the CMS Collaboration~\cite{Khachatryan:2015oea} does not separate the twist contribution from fluctuation of $v_n$ and contribution from fluctuation of $\Phi_{n}$. We propose a new observable involving correlations between four subevents in different $\eta$ intervals to separate the contribution from fluctuation of $v_n$ amplitude and contribution from fluctuation of $\Phi_{n}$. The CMS observable involves only the first moment of the $v_n$, which is extended to correlations between higher moments of the $v_n$ in two pseudorapidity ranges. We also propose to study the longitudinal correlations between harmonics of different order, e.g. between $v_2$ and $v_4$ in different $\eta$ intervals, to investigate how the mode-mixing effects evolves with rapidity. The expected strength of the signal from the initial state is estimated using a simple Glauber model. These observables together provide a framework for a comprehensive study of the longitudinal flow fluctuations in heavy ion collisions. 

\section{Moments of the longitudinal flow fluctuations}
We shall use the complex notation for harmonic flow. The azimuthal anisotropy of the particle production in an event is described by Fourier expansion of the underlying probability distribution $\emph{P}(\phi)$ in azimuthal angle $\phi$: 
\begin{eqnarray}
\label{eq:1}
\emph{P}(\phi) = \frac{1}{2\pi} \sum_{n=-\infty}^{\infty} {\bm v}_{n}e^{-in\phi},\;\;\; {\bm v}_n = v_n e^{in\Phi_n}\;,
\end{eqnarray}
where $v_n$ and $\Phi_n$ are magnitude and phase (or event plane), respectively. Due to the finite number of particles produced in each event, harmonic flow is estimated from the observed per-particle normalized flow vector ${\bm q}_n$:
\begin{eqnarray}
\label{eq:2}
{\bm q}_{n} \equiv \frac{\sum_i  w_ie^{in\phi_i}}{\sum_iw_i}\equiv q_{n}\;e^{in\Psi_n}
\end{eqnarray}
The sum runs over all $M$ particles in the event and $w_i$ is the weight assigned to $i^{\rm{th}}$ particle. 

Extending the observable used by CMS, we propose to study the longitudinal flow fluctuations using $\lr{{\bm q}_n^k (\eta_1) {\bm q}_n^{*k}(\eta_2)}$, the scalar product of the $k^{\rm {th}}$-moment~\cite{Bhalerao:2014xra} of flow vectors in two different $\eta$ intervals, averaged over events in a given centrality interval. This observable is effectively a $2k$-particle correlator between two subevent as defined in Ref.~\cite{Bilandzic:2013kga,jjia}~\footnote{It is denoted as $\lr{2k}_{ka|kb}$ in Ref.~\cite{jjia}, where ``a'' and ``b'' are subevents at $\eta_1$ and $\eta_2$, respectively}. Therefore one need to remove particle multiplets containing duplicated particle indexes. For example:
\begin{eqnarray}
\label{eq:b1}
{\bm q}_{n}^2\rightarrow ({\bm q}_{n}^2-\omega_{1}{\bm q}_{2n})/(1-\omega_1)\;,\;{\bm q}_{n}^3\rightarrow ({\bm q}_{n}^3-3\omega_{1}{\bm q}_{2n}{\bm q}_{n}+2\omega_{2}{\bm q}_{3n})/(1-3\omega_{1}+2\omega_{2})
\end{eqnarray}
where $\omega_k \equiv \frac{\sum_iw_i^{k+1}}{(\sum_iw_i)^{k+1}}$, and ${\bm q}_{kn} \equiv \frac{\sum_i  w_i^ke^{in\phi_i}}{\sum_iw_i^k}$. The correction terms are dropped from expressions for simplicity, but we assumed that they have been applied implicitly.

In the absence of non-flow correlation, the correlator can be expressed as:
\begin{eqnarray}
\label{eq:b2}
\lr{{\bm q}_n^k (\eta_1) {\bm q}_n^{*k}(\eta_2)}=\lr{{\bm v}_n^k (\eta_1) {\bm v}_n^{*k}(\eta_2)} = \lr{\left[v_n(\eta_1) v_n(\eta_2)\right]^k \cos kn(\Phi_n(\eta_1)-\Phi_n(\eta_2))}
\end{eqnarray}
where the statistical fluctuation drops out after averaging over events. The longitudinal flow decorrelations can be studied using the ratio in analogous to the CMS Collaboration~\cite{Khachatryan:2015oea}:
\begin{eqnarray}
\label{eq:b3}
r_{n|n;k}(\eta) = \frac{\lr{{\bm q}_n^k (-\eta) {\bm q}_n^{*k}(\etaref)}}{\lr{{\bm q}_n^k (\eta){\bm q}_n^{*k}(\etaref)}} = \frac{\lr{\left[v_n(-\eta) v_n(\etaref)\right]^k \cos kn(\Phi_n(-\eta)-\Phi_n(\etaref))}}{\lr{\left[v_n(\eta) v_n(\etaref)\right]^k \cos kn(\Phi_n(\eta)-\Phi_n(\etaref))}}
\end{eqnarray}
where the $\etaref$ is the reference pseudorapidity common to the numerator and denominator. This observable is sensitive to the forward-backward (FB) asymmetry of the $v_n$ magnitude as well as the twist of the event-plane angles between $\eta$ and $-\eta$.  Results on $r_{n|n;k}$ has been obtained by the CMS collaboration for $k=1$ and $n=2$--4, which is also simply denoted as $r_{n|n}$ in this note. However, measuring $r_{n|n;k}$ for $k>1$ can yield new information on how the $v_n$ asymmetry and event-plane twist fluctuate event by event.

In previous studies~\cite{Khachatryan:2015oea,Bozek:2015bna,Pang:2015zrq,Ma:2016fve,Schenke:2016ksl,Ke:2016jrd}, the $\etaref$ usually is chosen to be at large pseudorapidity, e.g. $\etaref>4$, while the pseudorapidity of ${\bm q}_n(-\eta)$ and ${\bm q}_n(\eta)$ is usually chosen to be close to mid-rapidity, $|\eta|<2.5$. This ensure a sizable pseudorapidity gap between ${\bm q}_n(\pm\eta)$ and ${\bm q}_n(\etaref)$ in Eq.~\ref{eq:b3}. The behavior of the flow longitudinal fluctuations can be discussed by extending the procedure of Ref.~\cite{Khachatryan:2015oea,Bozek:2015bna}. Assuming ${\bm v}_n$ in each event is a slowly varying function in $\eta$ near mid-rapidity,
\begin{eqnarray}
\label{eq:b4}
{\bm v}_n(\eta) \approx {\bm v}_n(0) (1+\alpha_n\eta)e^{i\beta_n\eta}\;,\;\;{\bm v}_n^k(0) {\bm v}_n^{*k}(\etaref) \equiv X_{n;k}(\etaref)-iY_{n;k}(\etaref),
\end{eqnarray}
then we have 
\begin{eqnarray}
\label{eq:b5}
\lr{{\bm q}_n^k (\eta){\bm q}_n^{*k}(\etaref)}&\approx& \lr{(1+k\eta\alpha_n)(X_{n;k}+k\eta\beta_nY_{n;k})}\approx\lr{X_{n;k}+k\eta\alpha_nX_{n;k}+k\eta\beta_nY_{n;k}}\\
&=&\lr{X_{n;k}}\left(1+k\eta\frac{\lr{\alpha_nX_{n;k}}}{\lr{X_{n;k}}}+k\eta\frac{\lr{\beta_nY_{n;k}}}{\lr{X_{n;k}}}\right)
\end{eqnarray}
and $r_{n|n;k}$ can be approximated by:
\begin{eqnarray}
\label{eq:b6}
&&r_{n|n;k}(\eta) = 1-2kF_{n;k}^{\rm{r}}\eta,\;\; F_{n;k}^{\rm{r}}\approx F_{n;k}^{\rm{asy}}+F_{n;k}^{\rm{twi}}\;,\;F_{n;k}^{\rm{asy}}\equiv \frac{\lr{\alpha_nX_{n;k}(\etaref)}}{\lr{X_{n;k}(\etaref)}}\;,F_{n;k}^{\rm{twi}} \equiv \frac{\lr{\beta_nY_{n;k}(\etaref)}}{\lr{X_{n;k}(\etaref)}},
\end{eqnarray}
where $F_{n;k}^{\rm{asy}}$ and $F_{n;k}^{\rm{twi}}$ represent the contribution from FB asymmetry and event-plane twist, respectively. Therefore a linear dependence on $\eta$ is expected near mid-rapidity with a slope of $F_{n;k}^{\rm{r}}$. Previous CMS measurement suggests that $F_{n;1}^{\rm{r}}$ has weak dependence on $\etaref$ for $\etaref>3$ at LHC. However recent calculations show that such dependence is expected at RHIC energy when $\etaref$ is reduced to $\etaref \approx 2.5$~\cite{Pang:2015zrq}. 

If $F_{n;k}^{\rm{asy}}$ and $F_{n;k}^{\rm{twi}}$ in Eq.~\ref{eq:b6} are also independent of $k$, i.e.
\begin{eqnarray}
\label{eq:b7}
F_{n;k}^{\rm{asy}}=F_{n;1}^{\rm{asy}}\equiv F_{n}^{\rm{asy}},\;\; F_{n;k}^{\rm{twi}} =F_{n;1}^{\rm{twi}}\equiv F_{n}^{\rm{twi}},
\end{eqnarray}
then one expects:
\begin{eqnarray}
\label{eq:b8}
r_{n|n;k}\approx r_{n|n;1}^k \;\mbox{or}\; r_{n|n}^k 
\end{eqnarray}
Deviation from this relation could provide insights on the detailed event-by-event structure in the longitudinal flow fluctuations.

\section{Separate the flow magnitude fluctuation and event-plane twist}
The observable $r_{n|n;k}$ does not separate the contribution of FB asymmetry of $v_n$ magnitude from event-plane twist effects (such separation can be partially done using event-shape engineering~\cite{Jia:2014ysa}). Therefore we introduce a new observable involving correlations of flow vectors in four $\eta$ intervals:
\begin{eqnarray}
\label{eq:c1}
\nonumber
R_{n,n|n,n}(\eta) &=& \frac{\lr{{\bm q}_n(-\etaref){\bm q}_n (-\eta){\bm q}_n^* (\eta) {\bm q}_n^*(\etaref)}}{\lr{{\bm q}_n(-\etaref){\bm q}_n^* (-\eta){\bm q}_n (\eta) {\bm q}_n^*(\etaref)}}\\\nonumber
&=&\frac{\lr{v_n(-\etaref)v_n(-\eta)v_n(\eta)v_n(\etaref)\cos n\left[\Phi_n(-\etaref)-\Phi_n(\etaref)+\;\;\Phi_n(-\eta)-\Phi_n(\eta)\;\;\right]}}{\lr{v_n(-\etaref)v_n(-\eta)v_n(\eta)v_n(\etaref)\cos n\left[\Phi_n(-\etaref)-\Phi_n(\etaref)-(\Phi_n(-\eta)-\Phi_n(\eta))\right]}}\\
\end{eqnarray}
The contribution from flow magnitude are identical between the numerator and denominators, and therefore this correlator is mainly sensitive to the event-plane twist effects. In fact, using the approximation Eq.~\ref{eq:b4} and assuming ${\bm q}_n(-\etaref){\bm q}_n^*(0) \approx {\bm q}_n^*(\etaref){\bm q}_n(0)$ we have:
\begin{eqnarray}
\label{eq:c2}
\nonumber
&&\hspace*{-1cm}\lr{{\bm q}_n(-\etaref){\bm q}_n (-\eta){\bm q}_n^* (\eta) {\bm q}_n^*(\etaref)}\approx \lr{{\bm q}_n(-\etaref){\bm q}_n^* (0){\bm q}_n (0) {\bm q}_n^*(\etaref)e^{-2i\beta_n\eta}}\\
\hspace*{1.5cm}&&\approx \lr{{\bm q}_n^2 (0) {\bm q}_n^{2*}(\etaref)e^{-2i\beta_n\eta}}=\lr{X_{n;2}}\left(1-2\eta\frac{\beta_nY_{n;2}}{\lr{X_{n;2}}}\right)
\end{eqnarray}
Therefore, the observable $R$ can be approximated as:
\begin{eqnarray}
\label{eq:c3}
R_{n,n|n,n} \approx 1-4\eta\frac{\lr{\beta_n Y_{n;2}(\etaref)}}{\lr{X_{n;2}(\etaref)}}\approx 1-4F_{n;2}^{\rm{twi}}\eta 
\end{eqnarray}
if $R_{n,n|n,n}$ is parameterized with a linear function similar to Eq.~\ref{eq:b6}, then:
\begin{eqnarray}
\label{eq:c4}
R_{n,n|n,n} =1-2F_n^{\rm{R}}\eta,\;\; F_n^{\rm{R}} \approx 2F_{n;2}^{\rm{twi}}
\end{eqnarray}
In this case, the observation of $F_n^{\rm{R}}>F_{n;2}^{\rm{r}}$ would imply that $F_{n;2}^{\rm{twi}}>F_{n;2}^{\rm{asy}}$, and vice versa. The contributions from FB asymmetry and event-plane twist can be estimated from $r_{n|n;2}$ and $R_{n,n|n,n}$:
\begin{eqnarray}
\label{eq:c5}
F_{n;2}^{\rm{twi}} = F_n^{\rm{R}}/2\;,\;\;\; F_{n;2}^{\rm{asy}} =  F_{n;2}^{\rm{r}}- F_n^{\rm{R}}/2\;,
\end{eqnarray}

\section{Longitudinal correlations of harmonics of different order}
The study of longitudinal flow fluctuations can also be extended to correlations between harmonics of different order.  The first interesting observable is $r_{2,3|2,3}$:
\begin{eqnarray}
\label{eq:d1}
r_{2,3|2,3}(\eta) = \frac{\lr{{\bm q}_2 (-\eta){\bm q}_2^{*}(\etaref){\bm q}_3 (-\eta){\bm q}_3^{*}(\etaref)}}{\lr{{\bm q}_2 (\eta){\bm q}_2^{*}(\etaref){\bm q}_3 (\eta){\bm q}_3^{*}(\etaref)}}= \frac{\lr{{\bm v}_2 (-\eta){\bm v}_2^{*}(\etaref){\bm v}_3 (-\eta){\bm v}_3^{*}(\etaref)}}{\lr{{\bm v}_2 (\eta){\bm v}_2^{*}(\etaref){\bm v}_3 (\eta){\bm v}_3^{*}(\etaref)}}
\end{eqnarray}
If the longitudinal fluctuations for $v_2$ and $v_3$ are independent of each other, one would expect $r_{2,3|2,3}\approx r_{2|2}r_{3|3}$. 

The other two interesting pseudorapidity correlation for $v_4$ and $v_5$ are:
\begin{eqnarray}
\label{eq:d2}
r_{2,2|4}(\eta) &=& \frac{\lr{{\bm q}_2^2 (-\eta) {\bm q}_4^{*}(\etaref)}+\lr{{\bm q}_2^2 (\etaref) {\bm q}_4^{*}(-\eta)}}{\lr{{\bm q}_2^2 (\eta){\bm q}_4^{*}(\etaref)}+\lr{{\bm q}_2^2 (\etaref){\bm q}_4^{*}(\eta)}}\\\label{eq:d3}
r_{2,3|5}(\eta) &=& \frac{\lr{{\bm q}_2(-\eta){\bm q}_3(-\eta) {\bm q}_5^{*}(\etaref) }+\lr{{\bm q}_2(\etaref){\bm q}_3(\etaref) {\bm q}_5^{*}(-\eta) }}{\lr{{\bm q}_2 (\eta){\bm q}_3 (\eta){\bm q}_5^{*}(\etaref)}+\lr{{\bm q}_2 (\etaref){\bm q}_3 (\etaref){\bm q}_5^{*}(\eta)}}
\end{eqnarray}
These observables are sensitive to the rapidity dependence of the correlations between event-plane of different order previously measured by ATLAS and ALICE experiments~\cite{Aad:2014fla,Aad:2015lwa,ALICE:2016kpq}, for example \mbox{$\lr{{\bm q}_2^2 (-\eta) {\bm q}_4^{*}(\etaref)} = \lr{v_2^2(-\eta) v_4(\etaref) \cos 4(\Phi_2(-\eta)-\Phi_4(\etaref))}$}. 

It is well-established that the ${\bm v}_4$ and ${\bm v}_5$ in heavy ion collisions contain linear contribution associated with initial geometry and mode-mixing contributions due to non-linear hydrodynamic response~\cite{Gardim:2011xv,Teaney:2012ke,Aad:2014fla,Aad:2015lwa,Qiu:2012uy}:
\begin{eqnarray}
\label{eq:d4}
{\bm v}_4 = {\bm v}_{4\rm{L}} +\beta_{2,2} {\bm v}_2^2\;,\;\;{\bm v}_5 = {\bm v}_{5\rm{L}}+\beta_{2,3} {\bm v}_2{\bm v}_3\;,
\end{eqnarray}
where the linear components are driven by the corresponding eccentricity ${\bm \epsilon}_n$ in the initial geometry, ${\bm v}_{4\rm{L}}\propto {\bm \epsilon}_4 $ and ${\bm v}_{5\rm{L}}\propto {\bm \epsilon}_5 $. Previous ATLAS measurements~\cite{Aad:2014fla,Aad:2015lwa} show that the linear component dominates in the most central collisions (0\%--5\% centrality interval), while the mode-mixing contribution dominates in other centrality interval. With this notation, Eqs.~\ref{eq:d2} and \ref{eq:d3} measure
\begin{eqnarray}
\label{eq:d5}
r_{2,2|4}(\eta) &=& \frac{\lr{{\bm v}_2^2 (-\eta) {\bm v}_{4\rm{L}}^{*}(\etaref)+{\bm v}_2^2 (\etaref) {\bm v}_{4\rm{L}}^{*}(-\eta)}+2\beta_{2,2}\lr{{\bm v}_2^2 (-\eta) {\bm v}_2^{*2}(\etaref)}}{\lr{{\bm v}_2^2 (\eta){\bm v}_{4\rm{L}}^{*}(\etaref)+{\bm v}_2^2 (\etaref){\bm v}_{4\rm{L}}^{*}(\eta)}+2\beta_{2,2}\lr{{\bm v}_2^2 (\eta) {\bm v}_2^{*2}(\etaref)}}\\\label{eq:d6}
r_{2,3|5}(\eta) &=& \frac{\lr{{\bm v}_2 (-\eta){\bm v}_3 (-\eta) {\bm v}_{5\rm{L}}^{*}(\etaref)+{\bm v}_2 (\etaref){\bm v}_3 (\etaref) {\bm v}_{5\rm{L}}^{*}(-\eta)}+2\beta_{2,3}\lr{{\bm v}_2 (-\eta){\bm v}_2^{*}(\etaref){\bm v}_3 (-\eta){\bm v}_3^{*}(\etaref)}}{\lr{{\bm v}_2 (\eta){\bm v}_3 (\eta) {\bm v}_{5\rm{L}}^{*}(\etaref)+{\bm v}_2 (\etaref){\bm v}_3 (\etaref) {\bm v}_{5\rm{L}}^{*}(\eta)}+2\beta_{2,3}\lr{{\bm v}_2 (\eta){\bm v}_2^{*}(\etaref){\bm v}_3 (\eta){\bm v}_3^{*}(\etaref)}}\\\nonumber
\end{eqnarray}
The correlation of the linear component of $v_4$ ($v_5$) with lower order harmonic  reflects mainly the correlation in the initial eccentricity:
\begin{eqnarray}
\label{eq:d7}
\lr{{\bm v}_{2}^2{\bm v}^*_{4\rm{L}}} \propto \lr{{\bm \epsilon}_{2}^2{\bm \epsilon}^*_{4}}\;,\;
\lr{{\bm v}_{2}{\bm v}_{3}{\bm v}^*_{5\rm{L}}} \propto \lr{{\bm \epsilon}_{2}{\bm \epsilon}_{3}{\bm \epsilon}^*_{5}}
\end{eqnarray}
These correlations are expected to be rather weak except in very peripheral collisions based on a Glauber model prediction~\cite{Jia:2012ma}. Furthermore the linear component of higher-order harmonics $v_4$ and $v_5$ suffer more viscous damping than the corresponding non-linear contributions $v_2^2$ and $v_2v_3$,  respectively. In fact, according to Refs.~\cite{Gubser:2010ui,Teaney:2012ke}, the correction for linear term scale as $\delta v_{n\rm{L}}\propto n^2\eta/s$, therefore 
\begin{eqnarray}
\nonumber
\delta v_{4\rm{L}}\propto -16\eta/s&,& \delta v_{4\rm{NL}}\propto \delta v^2_{2\rm{L}}\propto-8\eta/s, \\\label{eq:d7b}
\delta v_{5\rm{L}}\propto -25\eta/s&,& \delta v_{5\rm{NL}}\propto \delta v_{2\rm{L}}v_{3\rm{L}}\propto-13\eta/s.
\end{eqnarray}
Therefore the contributions in Eq.~\ref{eq:d7} are expected to be very small, and the values are expected to be similar between $r_{2,2|4}$ and $r_{2|2;2}$, and between $r_{2,3|5}$ and $r_{2,3|2,3}$. If this is true, then the pseudorapidity correlation for $v_4$ and $v_5$ can be approximated by:
\begin{eqnarray}
\label{eq:d8}
r_{4|4}(\eta) &\approx& \frac{\lr{{\bm v}_{4\rm{L}}(-\eta){\bm v}_{4\rm{L}}^{*}(\etaref)}+\beta_{2,2}^2\lr{{\bm v}_2^2 (-\eta) {\bm v}_2^{*2}(\etaref)}}{\lr{{\bm v}_{4\rm{L}}(\eta){\bm v}_{4\rm{L}}^{*}(\etaref)}+\beta_{2,2}^2\lr{{\bm v}_2^2 (\eta) {\bm v}_2^{*2}(\etaref)}}\\\label{eq:d9}
r_{5|5}(\eta) &\approx& \frac{\lr{{\bm v}_{5\rm{L}}(-\eta){\bm v}_{5\rm{L}}^{*}(\etaref)}+\beta_{2,3}^2\lr{{\bm v}_2 (-\eta){\bm v}_2^{*}(\etaref){\bm v}_3 (-\eta){\bm v}_3^{*}(\etaref)}}{\lr{{\bm v}_{5\rm{L}}(\eta){\bm v}_{5\rm{L}}^{*}(\etaref)}+\beta_{2,3}^2\lr{{\bm v}_2 (\eta){\bm v}_2^{*}(\etaref){\bm v}_3 (\eta){\bm v}_3^{*}(\etaref)}}\\\nonumber
\end{eqnarray}
Therefore $r_{4|4}$ and $r_{5|5}$ reflect the contributions from both linear component and non-linear component, but not the cross terms. Therefore comparing the three correlators, $r_{2|2;2}$, $r_{2,2|4}$ and $r_{4|4}$ allow us to separate the contributions of the linear and non-linear contribution to the longitudinal flow correlation of $v_4$. Similarly, comparing $r_{2,3|2,3}$, $r_{2,3|5}$ and $r_{5|5}$ allow us to separate the contributions of the linear and non-linear contribution to the longitudinal flow correlation of $v_5$.

For symmetric collisions, the numerator and denominator of the correlators Eqs.~\ref{eq:b3},~\ref{eq:c1},~\ref{eq:d1},~\ref{eq:d2} and \ref{eq:d3} can be symmetrized to double the statisics, by flipping the sign of the pseudorapidity,i.e, $\eta\rightarrow-\eta$ and  $\etaref\rightarrow-\etaref$. For example,  Eq.~\ref{eq:b3} becomes:
\begin{eqnarray}
\label{eq:d10}
r_{n|n;k}(\eta) = \frac{\lr{{\bm q}_n^k (-\eta) {\bm q}_n^{*k}(\etaref)+{\bm q}_n^k (\eta) {\bm q}_n^{*k}(-\etaref)}}{\lr{{\bm q}_n^k (\eta){\bm q}_n^{*k}(\etaref)+{\bm q}_n^k (-\eta){\bm q}_n^{*k}(-\etaref)}} 
\end{eqnarray}
and Eq.~\ref{eq:d5} becomes:
\begin{eqnarray}
\label{eq:d11}
r_{2,2|4}(\eta) &=& \frac{\lr{{\bm q}_2^2 (-\eta) {\bm q}_4^{*}(\etaref)+{\bm q}_2^2 (\etaref) {\bm q}_4^{*}(-\eta)}+\lr{{\bm q}_2^2 (\eta) {\bm q}_4^{*}(-\etaref)+{\bm q}_2^2 (-\etaref) {\bm q}_4^{*}(\eta)}}{\lr{{\bm q}_2^2 (\eta){\bm q}_4^{*}(\etaref)+{\bm q}_2^2 (\etaref){\bm q}_4^{*}(\eta)}+\lr{{\bm q}_2^2 (-\eta){\bm q}_4^{*}(-\etaref)+{\bm q}_2^2 (-\etaref){\bm q}_4^{*}(-\eta)}}
\end{eqnarray}

\section{Relation to initial eccentricity}
If the flow response is linear, then the $\eta$ asymmetry is controlled by the initial eccentricity of the forward and backward going nucleons ${\bm \epsilon}_n^{\rm F}$ and ${\bm \epsilon}_n^{\rm B}$, which has been confirmed by simulation based on AMPT model~\cite{Jia:2014ysa}. In this case, the total eccentricity of the initially produced fireball is expected to be a function of spatial rapidity $\eta$ as: 
\begin{eqnarray}
\label{eq:e1}\nonumber
{\bm \epsilon}_n(\eta) &=&{\bm \epsilon}_{n+}+f_n(\eta){\bm \epsilon}_{n-},\;\; {\bm \epsilon}_{n+} = \frac{{\bm \epsilon}_n^{\rm F}+{\bm \epsilon}_n^{\rm B}}{2}={\bm \epsilon}_n(0),\;\;{\bm \epsilon}_{n-} = \frac{{\bm \epsilon}_n^{\rm F}-{\bm \epsilon}_n^{\rm B}}{2}.
\end{eqnarray}
where the $f_n(\eta)$ is an odd function that controls th relative mixture of the eccentricity vectors from the forward and backward going nucleons: $f_n(\infty)=1$ and $f_n(-\infty)=-1$, and ${\bm \epsilon}_{n+}$ is the eccentricity calculated using all participants. The ${\bm \epsilon}_{n\pm}$ are constants within an event but fluctuate event to event.

In a simple linear response model, the $\eta$ dependence of the $r_{n|n;k}$ can be related to the initial eccentricity along spatial rapidity defined analogously to Eq.~\ref{eq:b3} (see also Ref.~\cite{Pang:2015zrq}):
\begin{eqnarray}
\label{eq:e2}
r^{\rm s}_{n|n;k}(\eta) = \frac{\lr{{\bm \epsilon}_n^k (-\eta) {\bm \epsilon}_n^{*k}(\etaref)}}{\lr{{\bm \epsilon}_n^k (\eta){\bm \epsilon}_n^{*k}(\etaref)}}
\end{eqnarray}
Assuming $f_n(\eta)$ in each event is a slowly varying function near mid-rapidity,
\begin{eqnarray}
\label{eq:e3}
{\bm \epsilon}_n(\eta) \approx {\bm \epsilon}_{n+} + {\bm \epsilon}_{n-} a_n\eta\;
\end{eqnarray}
then for symmetric collision systems, we have 
\begin{eqnarray}
\label{eq:e4}
\lr{{\bm \epsilon}_n^k (-\eta) {\bm \epsilon}_n^{*k}(\etaref)}&\approx& \lr{\epsilon_{n+}^{2k}} -k\eta \lr{a_nf(\etaref)k\epsilon_{n-}^{2}\epsilon_{n+}^{2k-2}}\\\label{eq:e5}
r^{\rm s}_{n|n;k}(\eta) &\approx& 1-2k\eta\frac{\lr{a_nf(\etaref)k\epsilon_{n-}^{2}\epsilon_{n+}^{2k-2}}}{\lr{\epsilon_{n+}^{2k}}}\approx 1-2k\eta\lr{a_nf(\etaref)}\frac{\lr{k\epsilon_{n-}^{2}\epsilon_{n+}^{2k-2}}}{\lr{\epsilon_{n+}^{2k}}}
\end{eqnarray}
where we only keep terms linear in $\eta$ or $f(\etaref)$,  and we use the fact ${\bm \epsilon}_{n-}$ changes sign when flipping ``F'' and ``B'', e.g. $\lr{{\bm \epsilon}_{n+}{\bm \epsilon}_{n-}^{*}}=0$. The last part of the Eq.~\ref{eq:e5} assumes that the fluctuations of $a_nf_n(\etaref)$ are independent of the fluctuations of ${\bm \epsilon}_{n}$.

Figure~\ref{fig:1} shows the Glauber model estimation of the value of $E_{n;k}\equiv\frac{\lr{k\epsilon_{n-}^{2}\epsilon_{n+}^{2k-2}}}{\lr{\epsilon_{n+}^{2k}}}$ for $n=2$ and 3 in Pb+Pb collisions. The value of $E_{2;k}$ increase with $k$, suggesting that the magnitude of the slope of $r^{\rm s}(\eta)$ scales faster than $k$  for $n=2$. On the other hand, the value of $E_{3;k}$ is nearly independent of $k$, implying that the slope of $r^{\rm s}(\eta)$ is proportional to $k$ for $n=3$.
\begin{figure}[h!]
\begin{center}
\includegraphics[width=0.9\linewidth]{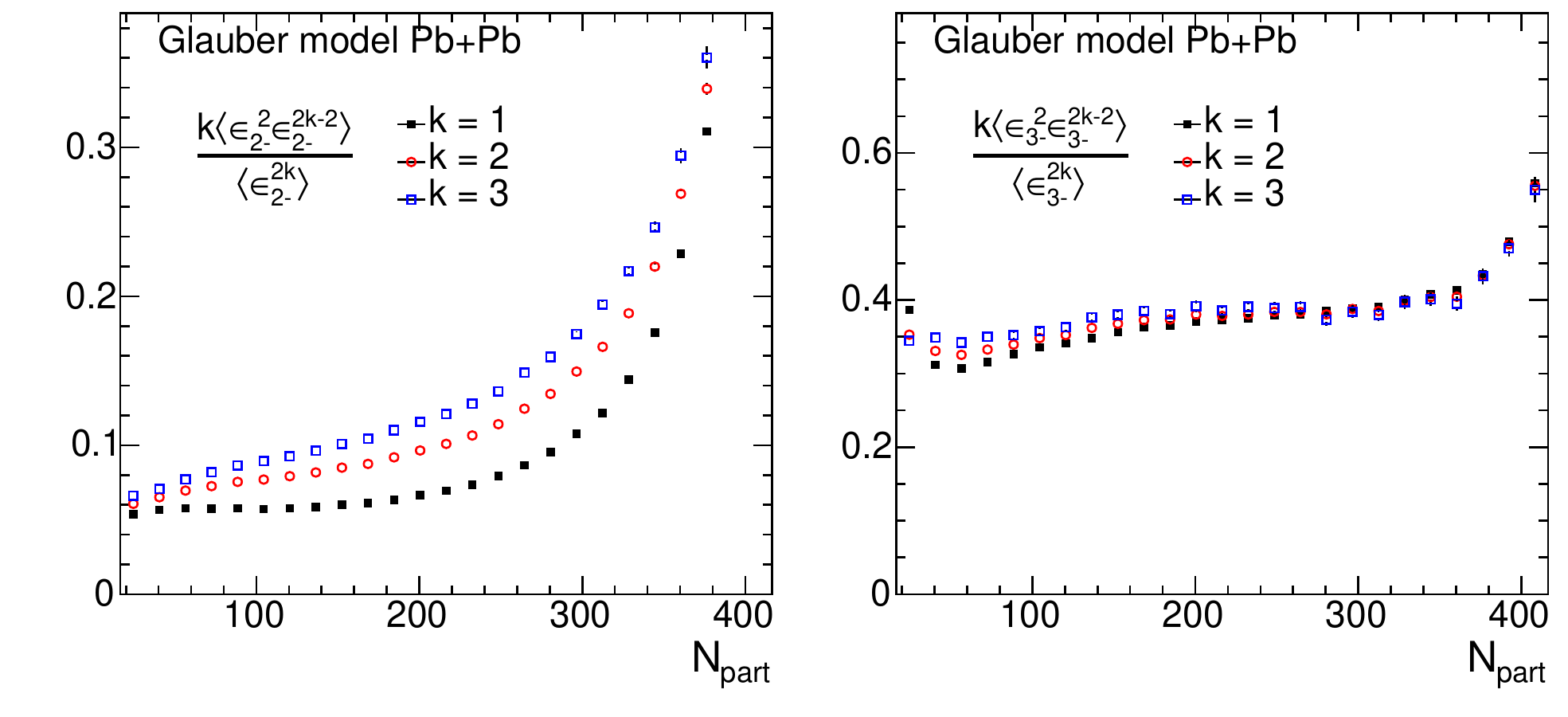}
\end{center}
\caption{\label{fig:1} The Glauber model estimation of $E_{n;k}\equiv\frac{\lr{k\epsilon_{n-}^{2}\epsilon_{n+}^{2k-2}}}{\lr{\epsilon_{n+}^{2k}}}$ for $n=2$ (left panel) and $n=3$ (right panel) in Pb+Pb collisions. They are shown for $k=1$, 2 and 3.}
\end{figure}

Let's now focus on separation of $r$ into the asymmetry and twist component, first we note that 
\begin{eqnarray}
\label{eq:e6}
&&16\epsilon_{n-}^{2}\epsilon_{n+}^{2} = \left((\epsilon_{n}^{\rm F})^2-(\epsilon_{n}^{\rm B})^2\right)^2 - \left({\bm \epsilon}_{n}^{\rm F}{\bm \epsilon}_{n}^{\rm B*}-{\bm \epsilon}_{n}^{\rm F*}{\bm \epsilon}_{n}^{\rm B}\right)^2 =A_1+A_2\\
&&A_1 \equiv \left((\epsilon_{n}^{\rm F})^2-(\epsilon_{n}^{\rm B})^2\right)^2\;,\;\;A_2\equiv -\left({\bm \epsilon}_{n}^{\rm F}{\bm \epsilon}_{n}^{\rm B*}-{\bm \epsilon}_{n}^{\rm F*}{\bm \epsilon}_{n}^{\rm B}\right)^2
\end{eqnarray}
The two parts correspond to the asymmetry component and twist component, respectively. Note that ${\bm \epsilon}_{n}^{\rm F}{\bm \epsilon}_{n}^{\rm B*}-{\bm \epsilon}_{n}^{\rm F*}{\bm \epsilon}_{n}^{\rm B}$ is a pure imaginary number and therefore $A_2>0$. With this we can rewrite $r^{\rm s}_{n|n;2}$ into asymmetry and twist components as:
\begin{eqnarray}
\label{eq:e7}
r^{\rm s}_{n|n;2}(\eta) &\approx& 1-4\eta (F_{n;2}^{\rm{s,asy}}+F_{n;2}^{\rm{s,twi}}),\; F_{n;2}^{\rm{s,asy}}= \frac{\lr{a_nf(\etaref)A_1}}{8\lr{\epsilon_{n+}^{4}}}\;,\;\;F_{n;2}^{\rm{s,twi}}=  \frac{\lr{a_nf(\etaref)A_2}}{8\lr{\epsilon_{n+}^{4}}}
\end{eqnarray}

Similarly the four particle correlator in Eq.~\ref{eq:c1} can be related to:
\begin{eqnarray}
\label{eq:e8}
R^{\rm s}_{n,n|n,n}=\frac{\lr{{\bm \epsilon}_n(-\etaref){\bm \epsilon}_n (-\eta){\bm \epsilon}_n^* (\eta) {\bm \epsilon}_n^*(\etaref)}}{\lr{{\bm \epsilon}_n(-\etaref){\bm \epsilon}_n^* (-\eta){\bm \epsilon}_n (\eta) {\bm \epsilon}_n^*(\etaref)}}&\approx& 1+2\eta\frac{\lr{a_nf_n(\etaref)({\bm \epsilon}_{n+}{\bm \epsilon}_{n-}^*-{\bm \epsilon}_{n+}^*{\bm \epsilon}_{n-})^2}}{\lr{\epsilon_{n+}^{4}}}\\
&=&1+2\eta\frac{\lr{a_nf_n(\etaref)({\bm \epsilon}_{n}^{\rm F}{\bm \epsilon}_{n}^{\rm B*}-{\bm \epsilon}_{n}^{\rm F*}{\bm \epsilon}_{n}^{\rm B})^2}}{4\lr{\epsilon_{n+}^{4}}}=1-4\eta F_{n;2}^{\rm{twi}}
\end{eqnarray}

Further assuming that the fluctuations of $a_nf_n(\etaref)$ are independent of the fluctuations of ${\bm \epsilon}_{n}$, we reach the following relation between the asymmetry and twist components:
\begin{eqnarray}
\label{eq:e9}
\frac{F_{n;2}^{\rm{s,twi}}}{F_{n;2}^{\rm{s,asy}}} \approx \frac{\lr{-\left({\bm \epsilon}_{n}^{\rm F}{\bm \epsilon}_{n}^{\rm B*}-{\bm \epsilon}_{n}^{\rm F*}{\bm \epsilon}_{n}^{\rm B}\right)^2}}{\lr{\left((\epsilon_{n}^{\rm F})^2-(\epsilon_{n}^{\rm B})^2\right)^2}}
\end{eqnarray}

Figure~\ref{fig:2} shows the Glauber model estimation of $\frac{F_{n;2}^{\rm{s,twi}}}{F_{n;2}^{\rm{s,asy}}}$ in Pb+Pb collisions. For $n=2$, the twist component is comparable to the asymmetry component in most central collisions, but is significantly larger towards more peripheral collisions. For $n=3$, the two components are comparable over the full centrality range, with twist component bing slightly larger in peripheral collisions.
\begin{figure}[h!]
\begin{center}
\includegraphics[width=0.6\linewidth]{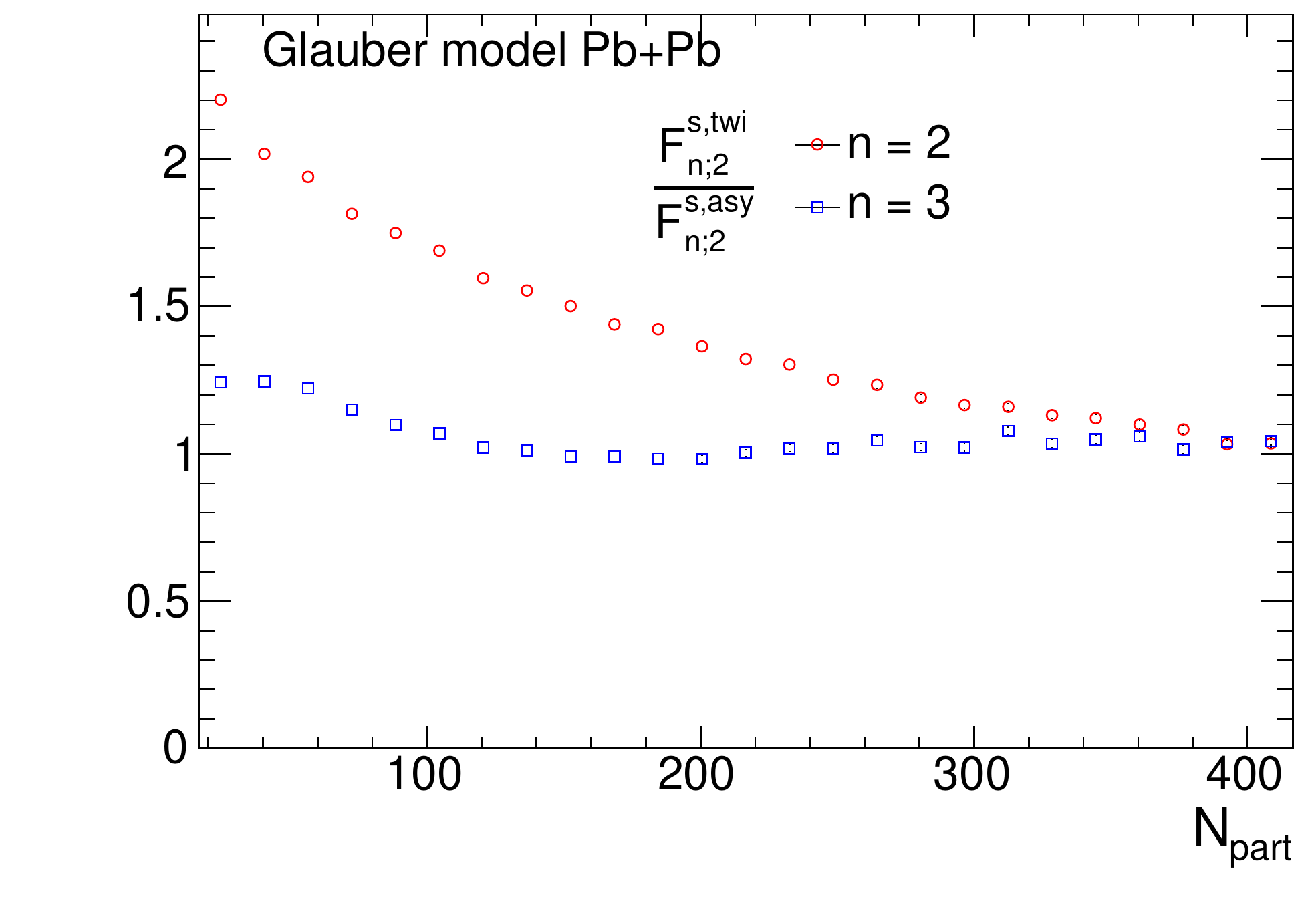}
\end{center}
\caption{\label{fig:2} The Glauber model estimation of the relative magnitudes of twist and asymmetry component of the longitudinal flow correlations for $n=2$ (open circles) and $n=3$ (open squares).}
\end{figure}

\section{Summary}
Motivated by the recent CMS measurement, we proposed several new correlators for studying the longitudinal flow correlations. These correlators are based on ratios of the moments of flow harmonics in two or more $\eta$-separated subevents, and can be used to infer the event-by-event flow fluctuations and mode-mixing effects in the longitudinal direction. Many of the correlators, especially those involving only $v_2$ or $v_3$, are sensitive to dynamics at early time, and therefore can be used to infer information of the eccentricity as a function of spatial rapidity. The latter directly constrains the initial condition that can be used by the state-of-art 3+1D event-by-event hydrodynamic models. These new correlators provide a framework for a comprehensive study of the longitudinal flow fluctuations in heavy ion collisions. 

A simple Glauber model that include forward-back asymmetry and $\eta$-dependent participant plane twist is used to predict the magnitude of the correlation as a function of the power of the moments, as well as the relative contributions from the asymmetry and twist. It is interesting to test if experimental data confirm these predictions.

J.~Jia and P. Huo acknowledge the support by NSF under grant number PHY-1613294. G.-L.M and M. Nie are supported by the National Natural Science Foundation of China under Grants No. 11522547, 11375251, and 11421505, and the Major State Basic Research Development Program in China under Grant No. 2014CB845404.
\bibliography{method2}{}
\bibliographystyle{apsrev4-1}
\end{document}